\documentclass[%
reprint,
superscriptaddress,
 amsmath,amssymb,
 aps,
floatfix,
]{revtex4-1}

\usepackage[dvipsnames,svgnames,x11names,hyperref]{xcolor}
\usepackage{siunitx}
\usepackage{tabularx}
\usepackage{placeins}

\usepackage{graphicx}% Include figure files
\usepackage{dcolumn}% Align table columns on decimal point
\usepackage{bm}% bold math
\usepackage{hyperref}% add hypertext capabilities
\usepackage[mathlines]{lineno}% Enable numbering of text and display math
\usepackage{mathrsfs}
\usepackage[normalem]{ulem}

\usepackage[margin=0.8in]{geometry}

\usepackage{braket}

\usepackage[normalem]{ulem}
\hypersetup{colorlinks=true, 
	linkcolor={blue!75!black!80!yellow},
	citecolor={blue!75!black!80!yellow}, 
	urlcolor=black %{blue!75!black!80!yellow}
	}
\makeatletter \renewcommand\@make@capt@title[2]{%
\@ifx@empty\float@link{\@firstofone}{\expandafter\href\expandafter{\float@link}}%
\sffamily{\textbf{#1}}\@caption@fignum@sep#2 }% \makeatother

\begin{document}
\preprint{APS/123-QED}

\title{Dipole-Coupled Defect Pairs as Deterministic Entangled Photon Pair Sources}

\author{Derek S. Wang}
%\email{derekwang@g.harvard.edu}
\thanks{These two authors contributed equally.}
\affiliation{Harvard John A. Paulson School of Engineering and Applied Sciences, Harvard University, Cambridge, MA 02138}
\author{Tom\'{a}\v{s} Neuman}
\thanks{These two authors contributed equally.}
\affiliation{Harvard John A. Paulson School of Engineering and Applied Sciences, Harvard University, Cambridge, MA 02138}
\author{Prineha Narang}
\email{prineha@seas.harvard.edu}
\affiliation{Harvard John A. Paulson School of Engineering and Applied Sciences, Harvard University, Cambridge, MA 02138}

\begin{abstract}
\noindent Scalable quantum systems require deterministic entangled photon pair sources. Here, we demonstrate a scheme that uses a dipole-coupled defect pair to deterministically emit polarization-entangled photon pairs. Based on this scheme, we predict spectroscopic signatures and quantify the entanglement with physically realizable system parameters. We describe how the Bell state fidelity and efficiency can be optimized by precisely tuning transition frequencies. A defect-based entangled photon pair source would offer numerous advantages including flexible on-chip photonic integration and tunable emission properties via external fields, electromagnetic environments, and defect selection.
\end{abstract}
\date{\today}

\maketitle

Non-classical states of light are important resources for quantum technologies, such as quantum information processing, networking, and metrology \cite{OBrien2009}. Entangled photon pairs, in particular, have applications in solid-state quantum repeaters, a crucial component of long-distance quantum networking that overcomes transmission loss by leveraging the effects of entanglement swapping and quantum teleportation \cite{Chen2016, Pan2012, Humphreys2018, Sangouard2011, Scarani2005, Zhao2010}. Despite the diverse applications for such non-classical states of light, methods for generating them deterministically remain limited. Currently, successful approaches are based on spontaneous parametric down-conversion \cite{Kwiat2001, Burnham1970} with high performance \cite{Lanco2006, Howell2004, Horn2012}. A major drawback of such methods is that the number of photon pairs generated follows a Poissonian distribution \cite{Waks2004}, rendering the pair generation efficiency too low for scalable quantum systems \cite{Pan2012}. While semiconductor quantum dots can deterministically emit entangled photon pairs via biexciton decay cascade, challenges including imperfections in synthesis \cite{Chen2016} motivate further exploration of potential materials systems. 

In this \textit{Letter}, we propose a scheme to deterministically generate entangled photon pairs from dipole-coupled defect pairs in solid-state materials. Defects in both 2D and 3D have wide applicability in quantum technologies, especially as quantum memories because they combine the favorable coherence and non-classical emission properties of isolated atoms \cite{Kurtsiefer2000, Grosso2017} with the scalability and stability of solid-state technologies \cite{Degen2017, Aharonovich2016, Atature2018, Childress2014}. A key breakthrough that highlights their applicability is the experimental demonstration of memory-enhanced quantum communication for quantum repeaters \cite{Bhaskar2019}. The ability to generate entangled photon pairs from defects would enable on-chip integration with quantum memories and emitters, minimizing the need to transduce photons from source to storage to emission in quantum technologies. 

\begin{figure}[tbhp]
\centering
\includegraphics[width=0.98\linewidth]{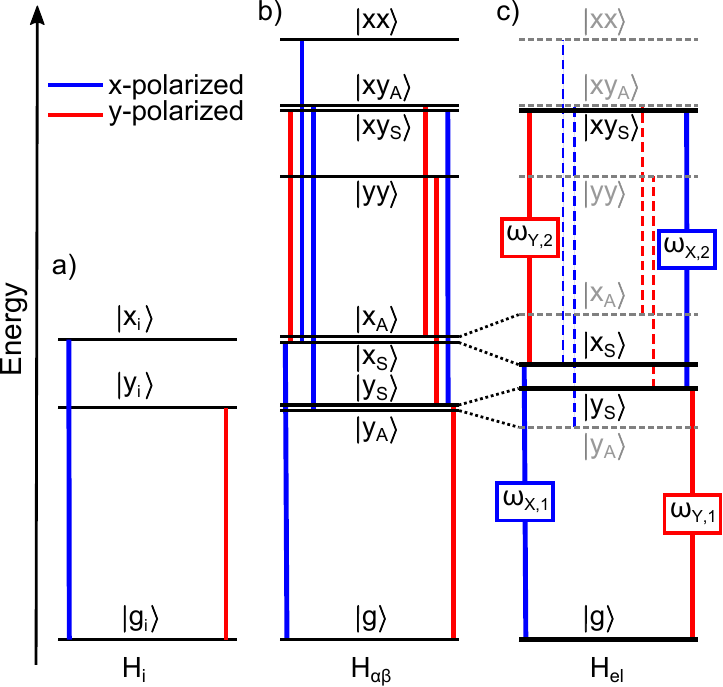}
\caption{Energy level diagrams and dipole-allowed transitions, where allowed $x$- and $y$-polarized transitions are in red and blue, respectively. \textbf{(a)} A single three-level defect. \textbf{(b)} Two distantly separated three-level defects such that dipole-coupling is negligible. \textbf{(c)} Two dipole-coupled, three-level defects. Bolded states and transitions (with transition frequencies $\omega_{X,1}$, $\omega_{X,2}$, $\omega_{Y,1}$ and $\omega_{Y,2}$) are accessible when the system is prepared in $|xy_{\rm S}\rangle$.}
\label{fig:diagram}
\end{figure}

We demonstrate how dipole-coupled defect pairs can generate polarization-entangled photon pairs. The system consists of two identical three-level systems denoted by $i\in\{\alpha,\beta\}$. Each three-level system consists of a ground state $|g_i\rangle$, excited state $|x_i\rangle$ with energy $\hbar\omega_x$ and transition dipole moment $\bm{d}_{x_i} =\langle x_i | {\rm e}\bm{r} | g_i \rangle=d_{x_i} \hat{x}$, and excited state $|y_i\rangle$ with energy $\hbar\omega_y$ and  transition dipole moment $\bm{d}_{y_i}=\langle y_i | {\rm e}\bm{r} | g_i \rangle=d_{y_i}\hat{y}$, where $\bm{r}$ is the position operator and ${\rm e}$ is the electron charge. The energy level diagram and dipole-allowed transitions are plotted in Fig. \ref{fig:diagram}(a). The Hamiltonian $H_i$ of each isolated three-level system can be written as $H_i = \hbar\omega_x |x_i \rangle \langle x_i| + \hbar\omega_y | y_i \rangle \langle y_i|$.

When defects $\alpha$ and $\beta$ at positions $\bm{r}_\alpha$ and $\bm{r}_\beta$, respectively, are brought close and couple via electric dipole interactions, the total electronic Hamiltonian $H_{\rm el}$ can be written in the product space of the two three-level systems as

\begin{align} \label{eq:hamiltonian}
    H_{\rm el} = {H_{\alpha\beta}} + H_{\rm dip}, 
\end{align}
where $H_{\alpha\beta}=H_\alpha+H_\beta$, and the dipole-coupling Hamiltonian $H_{\rm dip}$, in the rotating wave approximation (RWA) where we have dropped double (de-)excitations, is given by
%% TOMAS
\begin{align}
    H_{\rm dip}=\sum_{pq\in \{x,y\}}J_{pq}(|g p\rangle \langle q g| +|q g\rangle \langle g p|),
\end{align}
where $|rs\rangle\equiv |r_\alpha\rangle |s_\beta\rangle$ with $r,s \in \{g, x, y\}$, and transition dipole moments are real. Although we assume defect states do not have permanent dipole moments, we can include easily their interactions as diagonal terms in the single defect subspace. We also assume the orbitals of neighboring defects do not hybridize in the interdefect ranges considered of a few to tens of nanometers because defect orbitals can be localized within a few \AA~\cite{Harris2019, Hayee2020, Trusheim2019, McDougall2017}. The dipole interaction energy $J_{pq}$ is
\begin{align}
    J_{pq} = \frac{|\bm{d}_{p_\alpha} ||\bm{d}_{q_\beta} |}{4\pi\epsilon_0\epsilon_r|\bm{r}_\alpha-\bm{r}_\beta|^3}\left[\bm{e}_{p_\alpha}\cdot\bm{e}_{q_\beta}-3(\bm{e}_{p_\alpha}\cdot\bm{n})(\bm{e}_{q_\beta}\cdot\bm{n})\right],
\end{align}
where $\epsilon_r$ is the relative permittivity of the host material, $\bm{e}_{s_i}$ is the unit vector of the dipole moment $\bm{d}_{s_i}$, and $\bm{n}$ is the unit vector of $\bm{r}_\alpha-\bm{r}_\beta$. Given that transition dipole moments of $\sim$1 e\AA~have been experimentally observed, we estimate that emitters spaced a few nm apart can have dipole interaction energies on the order of tens of \si{\micro\electronvolt}.

Assuming $\bm{n}$ lies on the $x$-axis, $d_x\equiv d_{x_\alpha}=d_{x_{\beta}}$, and $d_y \equiv d_{y_\alpha}=d_{y_{\beta}}$, $H_{\rm el}$ can be diagonalized to produce nine eigenstates with eigenenergies listed in Table \ref{tab:eigen}. The subscripts ``A" and ``S" stand for ``anti-symmetric" and ``symmetric" combinations, respectively. The energy diagram of the eigenstates of $H_{\alpha\beta}$ and $H_{\rm el}$ and their dipole-allowed transitions, derived from the dipole operator $\bm{d}$ listed in Appendix \ref{sec:dipole}, are plotted in Fig. \ref{fig:diagram}(b)-(c). Notably, direct transitions between symmetric and anti-symmetric states are dipole-forbidden. From the energy diagram corresponding to $H_{\rm el}$, we see that a polarization-entangled photon pair can be emitted when the system is prepared in $|xy_{\rm S}\rangle$ and irreversibly decays. 

\begin{table}[h!]
\centering
\begin{tabular}{ ||c|c|c||} 
\hline
& Eigenstate & Eigenenergy \\
 \hline
1 & $|g\rangle \equiv |g g\rangle$ & $\hbar\omega_g=0$ \\
2 & $|y_{\rm A}\rangle \equiv \frac{1}{\sqrt{2}}(|gy\rangle - |yg\rangle)$ & $\hbar\omega_{y_{\rm A}}=\hbar\omega_y-J_{yy}$ \\
3 & $|y_{\rm S}\rangle \equiv \frac{1}{\sqrt{2}}(|gy\rangle + |yg\rangle)$ & $\hbar\omega_{y_{\rm S}}=\hbar\omega_y+J_{yy}$\\
4 & $|x_{\rm S}\rangle \equiv \frac{1}{\sqrt{2}}(|gx\rangle + |xg\rangle)$ & $\hbar\omega_{x_{\rm S}}=\hbar\omega_x-J_{xx}$\\
5 & $|x_{\rm A}\rangle \equiv \frac{1}{\sqrt{2}}(|gx\rangle - |xg\rangle$ & $\hbar\omega_{x_{\rm A}}=\hbar\omega_x+J_{xx}$\\
6 & $ |yy\rangle$ & $\hbar\omega_{yy}=2\hbar\omega_y$\\
7 & $|xy_{\rm S}\rangle \equiv \frac{1}{\sqrt{2}}(|xy\rangle + |yx\rangle)$ & $\hbar\omega_{xy_{\rm S}}=\hbar(\omega_x+\omega_y)$\\ 
8 & $|xy_{\rm A}\rangle \equiv \frac{1}{\sqrt{2}}(|xy\rangle - |yx\rangle)$ & $\hbar\omega_{xy_{\rm A}}=\hbar\omega_{xy_{\rm S}}$\\
9 & $|xx\rangle$ & $\hbar\omega_{xx}=2\hbar\omega_{x}$ \\
\hline
\end{tabular}
\caption{Eigenstates and eigenenergies of $H_{\rm el}$.}
\label{tab:eigen}
\end{table}

We calculate emission spectra into free space by coupling the defect system initially prepared in $|xy_{\rm S}\rangle$ to an unexcited continuum of photon modes and solving the time-dependent Schr\"odinger equation under the Weisskopf-Wigner approximation \cite{Weisskopf1930}, similarly to the approach introduced in Ref. \cite{Pathak2009generation}. A potential pumping scheme is described in Appendix \ref{sec:pump}.

The total Hamiltonian $H$ of the coupled defect-photon system is
\begin{align}
    H = H_{\rm el} + H_{\rm ph} + H_{\rm el-ph}.
\end{align}
The photonic Hamiltonian $H_{\rm ph}$ is $H_{\rm ph}=\sum_{jl} \hbar   \omega_j a^\dagger_{jl} a_{jl}$, where $a_{jl}$ ($a_{jl}^\dagger$) are annihilation (creation) operators of the $j$th mode in the electromagnetic vacuum of free space with polarization $l \in \{X,Y\}$ and energy $\hbar\omega_j$. 
In $H_{\rm ph}$, we have dropped the zero-point contribution with no loss of generality. 
The electron-photon coupling Hamiltonian in the RWA and dipole approximation is $H_{\rm el-ph}=-\sum_{opjl} \bm{\mathcal{E}}_{jl} \cdot \bm{d}_{op} |o \rangle \langle p | a_{jl}^\dagger+ {\rm H.c.}$, where $\bm{\mathcal{E}}_{jl}$ is the electric field with magnitude $\mathcal{E}$ in the $l$ direction that we assume to be constant for all $j$, and $\bm{d}_{op}=\langle o|{\rm e}\bm{r}|p\rangle$ with $|o\rangle$ and $|p\rangle$ being quantum states of the combined two-emitter system.

The \textit{ansatz} for a general electron-photon wave function, noting that for a system prepared in $|xy_{\rm S}\rangle$ there can be a maximum of two excitations distributed among the electronic and photonic states, is
\begin{multline} \label{eq:generalwfn}
    |\Psi(t)\rangle = \sum_{jk} c_{jk}^g|g\rangle a_{jX}^\dagger a_{kY}^\dagger |{\rm vac}\rangle + \sum_{j} c_{j}^{x_{\rm S}}|x_{\rm S}\rangle a_{jY}^\dagger|{\rm vac}\rangle \\ + \sum_{j} c_{j}^{y_{\rm S}}|y_{\rm S}\rangle a_{jX}^\dagger |{\rm vac}\rangle + c^{xy_{\rm S}} |xy_{\rm S}\rangle|{\rm vac}\rangle,
\end{multline}
where $j$ and $k$ are indices for the continuum of photon modes and $|{\rm vac}\rangle$ is the photon vacuum state, and $c_{jk}^g$, $c_{j}^{x_{\rm S}}$, $c_{j}^{y_{\rm S}}$ and $c^{xy_{\rm S}}$ are time-dependent amplitudes. We have dropped all anti-symmetric, $|yy\rangle$, and $|xx\rangle$ terms because the defect system is initially prepared in $|xy_{\rm S}\rangle$.

We solve the time-dependent Schr\"odinger equation under the Weisskopf-Wigner approximation to find the the final state of the electron-photon system \cite{Pathak2009generation}:
\begin{equation} \label{eq:steadystatewfn}
    |\Psi(\infty)\rangle = \sum_{jk} c_{jk}^g(\infty) |g\rangle a_{jX}^\dagger a_{kY}^\dagger |{\rm vac}\rangle,
\end{equation}
where
\begin{multline} \label{eq:sscoefficient}
    c_{jk}^g(\infty)=\frac{\frac{-\Omega_{g,x_{\rm S}}\Omega_{x_{\rm S},xy_{\rm S}}}{i\omega_{x_{\rm S}}-i\omega_j+\gamma_{g,x_{\rm S}}}+\frac{-\Omega_{g,y_{\rm S}}\Omega_{y_{\rm S},xy_{\rm S}}}{i\omega_{y_{\rm S}}-i\omega_k+\gamma_{g,y_{\rm S}}}}{i(\omega_{xy_{\rm S}}-\omega_j-\omega_k)+\gamma_{x_{\rm S},xy_{\rm S}}+\gamma_{y_{\rm S},xy_{\rm S}}},
\end{multline}
and $\Omega_{op}=-\mathcal{E}|\bm{d}_{op}|/\hbar$, $\gamma_{op}=\mathcal{E} ^2|\bm{d}_{op}|^2/\Delta$, and $\Delta$ is the frequency spacing. Further details on obtaining Eq. \eqref{eq:sscoefficient} are in Appendix \ref{sec:wwapproximation}.

\begin{figure}[!ht]
\centering
\includegraphics[width=1.0\linewidth]{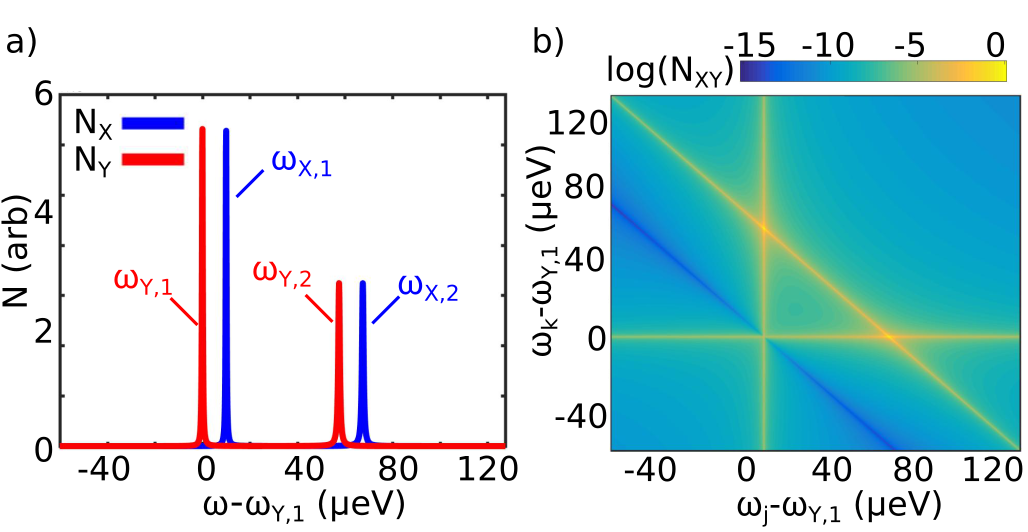}
\caption{Spectra of emitted polarization-entangled photon pair. \textbf{(a)} The single-photon spectra $N_X(\omega_j)$ and $N_Y(\omega_k)$ corresponding to $x$- and $y$-polarized photons, respectively, and \textbf{(b)} the cross-correlation function $N_{XY}(\omega_j,\omega_k)$. Based on experimentally observed ranges of parameters, we set $\omega_{y_{\rm S}}=2$ eV, $\omega_{x_{\rm S}}=\omega_{y_{\rm S}}+10$\,\si{\micro\electronvolt}, $d_x=d_y=1$ e\AA, $|\bm{r}_\alpha-\bm{r}_\beta|=5$ nm, $\epsilon_r=2$, and $\gamma_{g,y_{\rm S}}=0.2$\,\si{\micro\electronvolt}.}
\label{fig:ideal}
\end{figure}

\begin{figure*}[!htb]
\centering
\includegraphics[width=1.0\linewidth]{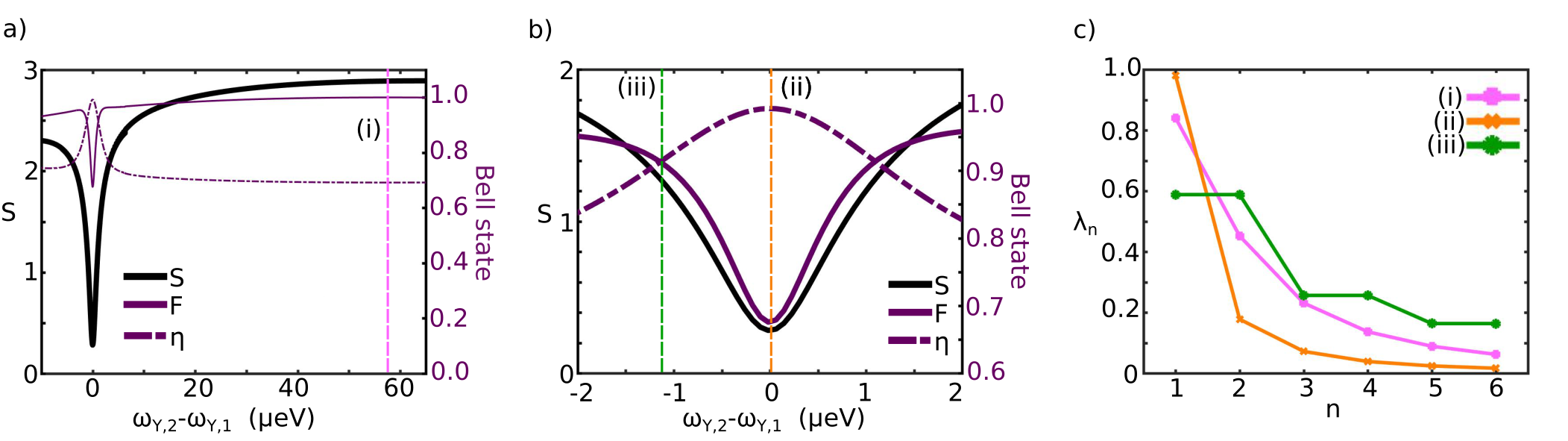}
\caption{Entanglement optimization. \textbf{(a)} Entanglement entropy $S$, Bell state efficiency $\eta$, and Bell state fidelity $\mathcal{F}$ for varying $\omega_{Y,2}-\omega_{Y,1}=\omega_{X,2}-\omega_{X,1}$, effected by changing $d_x$. The pink line (i) corresponds to the conditions in Fig. \ref{fig:ideal}. \textbf{(b)} Magnified near $\omega_{Y,2}-\omega_{Y,1}=0$. Both $S$ and $\mathcal{F}$ are minimized at (ii), and both $\eta$ and $\mathcal{F}>0.90$ at (iii). \textbf{(c)} Singular values (wave function coefficients) of entangled photon pairs corresponding to conditions marked by (i), (ii), and (iii) in Fig. \ref{fig:sweep}(a)-(b).}
\label{fig:sweep}
\end{figure*}

We explore the physical parameters that result in photon pair entanglement. First, we calculate spectra for a photon pair emitted by a dipole-coupled defect pair and note spectral signatures of entanglement. We optimize the Bell state fidelity by tuning transition frequencies. These changes can be implemented by appropriate selection of a defect system or applying external fields.

The emission cascade caused by the radiative decay of the optically excitable $|xy_{\rm S}\rangle$ state of the composite emitter-emitter system results in the emission of $x-$ and $y-$polarized photons whose number spectra are generally distinct, as we show in Fig. \ref{fig:ideal}(a) for parameters given in the figure caption. We calculate the number spectra, or the probability of finding an $x-$polarized ($y-$polarized) photon with frequency $\omega_j$ [$\omega_k$], as $N_X(\omega_j)=\sum_{j} |c^{\rm g}_{jk}|^2$ [$N_Y(\omega_k)=\sum_{j} |c^{\rm g}_{jk}|^2$]. While the $x-$polarized photon spectrum $N_X(\omega_j)$ (blue curve) peaks around the frequencies $\omega_{X,1}$ and $\omega_{X,2}$, the maxima of the $y-$polarized spectrum are found at $\omega_{Y,1}$ and $\omega_{Y,2}$, corresponding to the respective transitions in the two-photon cascade depicted in Fig. \ref{fig:diagram}(c) as blue and red lines.

The emitted $x-$ and $y-$polarized photons of different frequencies exhibit nontrivial correlations. We plot in Fig. \ref{fig:ideal}(a) the cross-correlation function $N_{XY}(\omega_j, \omega_k)=|c_{jk}^g|^2$ measuring the probability to simultaneously detect an $x$-polarized photon of frequency $\omega_j$ and an $y$-polarized photon of frequency $\omega_k$. The cross-correlation function features local maxima at two points. When an $x$-polarized photon is detected with frequency $\omega_{X,1}$, the $y$-polarized photon is most likely detected with frequency $\omega_{Y, 2}$ [i.e. $N_{XY}(\omega_{X,1}, \omega_{Y,2})$ is a maximum], and when an $x$-polarized photon is detected with frequency $\omega_{X,2}$, the probability of simultaneously finding an $y$-polarized photon peaks for frequency $\omega_{Y,1}$.
%Importantly, the cross-probabilities $N_{XY}(\omega_{X,1}, \omega_{Y,1})$ and $N_{XY}(\omega_{X,2}, \omega_{Y,2})$ practically vanish. 
This correlated behavior for a pure state is an intuitive signature of bipartite entanglement. 

We consider two metrics to rigorously quantify the entanglement of emitted photon pairs. The first metric is the entanglement entropy $S$ \cite{Parker2000, Law2000, Huang1993}:
\begin{align}
    S=-\sum_n |\lambda_n|^2 {\rm log}_2 |\lambda_n|^2.
\end{align}
We find the singular values $\lambda_n$ by Schmidt decomposition of the photonic portion $|\Psi_{\rm ph}\rangle$ of the final state in Eq. \eqref{eq:steadystatewfn}:
\begin{align} \label{eq:schmidtdecompose}
        |\Psi_{\rm ph}\rangle = \sum_{n} \lambda_n b_{nX}^\dagger c_{nY}^\dagger |{\rm vac}\rangle,
\end{align}
where the creation operators $b_{nX}^\dagger= \sum_j \psi_{nj}a_{jX}^\dagger$ and $c_{nY}^\dagger=\sum_k \phi_{nk}a_{kY}^\dagger$ in the Schmidt basis, $\lambda_n$ represent wave function coefficients in decreasing order with $n$, and $\psi_{nj}$ and $\phi_{nk}$ are the eigenfunctions of $c_{jk}^g$. The entanglement entropy is zero if the state is factorizable and greater than zero for an entangled state.

In protocols based on entanglement, it is often convenient to work directly with Bell states, so the second and third metrics we consider are the Bell state efficiency $\eta$ and fidelity $\mathcal{F}$, where the Bell state $|\Psi^+\rangle=\frac{1}{\sqrt{2}}(|10\rangle +|01\rangle )$ in the logical basis. To write $|\Psi_{\rm ph}\rangle$ in the logical basis, we assign the Schmidt states defined by the two pairs of $b_{nX}^\dagger$ and $c_{nY}^\dagger$ with the highest $\lambda_n$ to $|10\rangle $ and $|01\rangle $:
\begin{equation}
        |\Psi_{\rm ph}\rangle = \lambda_0 |10\rangle  + \lambda_1 |01\rangle  + \sum_{n=2} \lambda_n b_{nX}^\dagger c_{nY}^\dagger |{\rm vac}\rangle.
\end{equation}
We trace out all states where $n\geq 2$ to write the reduced density matrix $\rho_{\rm R}$ as
\begin{align}
    \rho_{\rm R} = (\lambda_0^2+\lambda_1^2)|\psi \rangle \langle \psi | + \sum_{n=2} \lambda_n ^2 |00\rangle  \langle 00 |,
\end{align}
where $|\psi\rangle = 1/\sqrt{\lambda_0^2+\lambda_1^2}(\lambda_0 |10\rangle +\lambda_1 |01\rangle )$.
The efficiency $\eta$ of collecting $|10\rangle $ and $|01\rangle $ is
\begin{align}
    \eta &= \lambda_0^2+\lambda_1^2,
\end{align}
and the Bell state fidelity $\mathcal{F}=|\langle \Psi^+ | \psi \rangle |^2$ is
\begin{align}
    \mathcal{F}=\frac{1}{2}\frac{(\lambda_0+\lambda_1)^2}{\lambda_0^2+\lambda_1^2}.
\end{align}

In Fig.\,\ref{fig:sweep} we show how the entanglement can be optimized by tuning defect parameters. In Fig.\,\ref{fig:sweep}(a), we sweep $d_x$ while holding all other physical parameters described in Fig.\,\ref{fig:ideal} constant. As a result, $\omega_{Y,2}$ [$\omega_{X,2}$] shifts relative to $\omega_{Y,1}$ [$\omega_{X,1}$], modulating the distance between peaks of the single-photon spectrum of a given polarization. Notably, for the exact conditions plotted in Fig. \ref{fig:ideal}, $d_x=d_y$, $\mathcal{F}$ is nearly 1 while $\eta=0.69$. In Fig. \ref{fig:sweep}(b), we zoom into the region around $\omega_{Y,2}= \omega_{Y,1}$, corresponding to $d_x=\frac{1}{\sqrt{2}}d_y$. Here we observe a minimum in $S$ and $\mathcal{F}$ and a maximum in $\eta$. The entanglement entropy drops here because the frequency of a photon with a given polarization emitted by one of the two decay paths is the same as the photon with a given polarization emitted via the other decay path, so photon pairs emitted by either of the two decay paths are identical. The finite linewidth of the emissions, however, permits entanglement among photon modes within this peak, so the entanglement entropy does not bottom out at 0.

$\mathcal{F}$ and $\eta$ of the emitted photon pair change in opposite directions surrounding the minimum of $\mathcal{F}$ and $S$. To understand the origin of this observation, in Fig.\,\ref{fig:sweep}(c) we plot the first few Schmidt coefficients $\lambda_n$ when: (i) $d_x=d_y$ corresponding to the state analyzed in Fig.\,\ref{fig:ideal}, (ii) $S$ and $\mathcal{F}$ are minimized, and (iii) both $\eta$ and $\mathcal{F}>0.90$. In (i), we see that $\lambda_n$ come in pairs, meaning that this state is a superposition of high-fidelity polarization-entangled Bell states in different bases. In (ii), where $S$ and $\mathcal{F}$ are minimized, $\lambda_n$ decays more quickly than in (i). Nearly all of the population is concentrated in the first state, so there are fewer entangled states, lowering $S$. A balance is achieved in (iii) where probability density is concentrated within the first two pairs of entangled states, but $\lambda_1 \neq \lambda_2$. Thus, by tuning the transition frequencies, we can optimize for $\mathcal{F}$ or $\eta$.
The entanglement measures are robust to changes in $\omega_{X,1}-\omega_{Y,1}$ as shown in Fig. \ref{fig:wxSwyS}. Finally, we note that the emitted photon pairs can undergo entanglement distillation to further enhance the Bell state fidelity \cite{Zhao, Bose, Parker2000, Bennett199653, Krastanov2019}.

The present study provides the theoretical basis for a defect-based, deterministic entangled photon pair source whose emission can be tailored chemically or externally and can be integrated on-chip for a variety of quantum technologies. Specifically, we dipole-couple two three-level defect systems, each with excited states with orthogonal transition dipole moments, to form a composite defect system. When the composite defect system is excited to a symmetric doubly excited state and subsequently de-excites in a radiative cascade, two entangled photons are emitted. 
We find that the entanglement measures of the emitted photons are robust to relative differences in frequency between the intermediate states. 
Importantly, the Bell state fidelity $\mathcal{F}$ and efficiency $\eta$ can be optimized by e.g. tuning the defect transition dipole moments.

The proposed scheme requires a defect system with two orthogonally polarized excited states, and we expect it to be possible for this condition to be fulfilled. The chemical selection space of defect systems is vast, as the chemical identity of the defect and surrounding matrix can be permuted to discover the appropriate system for a specific application \cite{Narang2019}. In addition, because the double excitation is delocalized as single excitations on two sites, rather than concentrated on a single site in the biexciton decay cascade, and because accurately computing multiply excited states remains a significant challenge \cite{Loos2019}, the present scheme is more amenable to computational searches of defect system candidates. As is the case in semiconductor quantum dots \cite{Pathak2009generation, Poddubny2012, Young2006, Trivedi2020}, system imperfections can be  modulated by coupling defects to external fields, including electric, magnetic, and strain, as well as to waveguides and cavity environments. These effects have been studied extensively in defect systems \cite{Rogers2008, Momenzadeh2015, Faraon2012, Chakraborty2019, Zhang2018, Machielse2019}, thereby enabling near-term experimental observations of the present proposal.

Accounting for dephasing and losses due to phonons or many-body effects is a natural extension of the present model. Further studies could also explore how external fields and sculpted electromagnetic environments could boost the entanglement and fidelity to improve the practical applicability of the proposed scheme. In addition, taking full advantage of the entanglement entropy of entangled photon states in the continuum beyond Bell states warrants further investigation.

\section*{Acknowledgements}
We acknowledge fruitful discussions with Stefan Krastanov, Matthew Trusheim, and Dirk Englund. This work was supported by the Department of Energy `Photonics at Thermodynamic Limits’ Energy Frontier Research Center under grant DE-SC0019140. T.N. is partially supported by the U.S. Department of Energy, Office of Science, Basic Energy Sciences (BES), Materials Sciences and Engineering Division under FWP ERKCK47 `Understanding and Controlling Entangled and Correlated Quantum States in Confined Solid-state Systems Created via Atomic Scale Manipulation'. D.W. is an NSF Graduate Research Fellow. P.N. is a Moore Inventor Fellow through Grant GBMF8048 from the Gordon and Betty Moore Foundation. D.W. and T.N contributed equally to this work.

\appendix
\section{Dipole operator} \label{sec:dipole}
We explicitly write the dipole operator in the eigenbasis of the total electronic Hamiltonian $H_{\rm el}$.

\begin{table}[h!]
\centering
\begin{tabular}{ ||c|c|c||} 
\hline
Initial & Final & $\bm{d}$ \\
 \hline
$|g\rangle$ & $|x_{\rm S} \rangle$ & $\sqrt{2}d_x\hat{x}$ \\
$|g\rangle$ & $|y_{\rm S} \rangle$ & $\sqrt{2}d_y\hat{y}$ \\
$|x_{\rm S}\rangle$ & $|xy_{\rm S}\rangle$ & $d_y\hat{y}$ \\
$|y_{\rm S} \rangle$ & $|xy_{\rm S}\rangle$ & $d_x\hat{x}$ \\
$|x_{\rm S} \rangle$ & $|xx\rangle$ & $\sqrt{2} d_x\hat{x}$ \\
$|y_{\rm S} \rangle$ & $|yy\rangle$ & $\sqrt{2} d_y\hat{y}$ \\
$|x_{\rm A}\rangle$ & $|xy_{\rm A}\rangle$ & $d_y\hat{y}$ \\
$|y_{\rm A}\rangle$ & $|xy_{\rm A}\rangle$ & $d_x\hat{x}$ \\
\hline
\end{tabular}
\caption{The dipole operator $\bm{d}$ in the eigenbasis.}
\label{tab:dipole}
\end{table}

\section{Initialization of the system} \label{sec:pump}

To generate the entangled phonon pairs it is first necessary to efficiently excite the coupled-defect system to the symmetrical state $|xy_{\rm S}\rangle$. Here we describe an example pumping scheme involving two-photon absorption. We consider a general scenario where the transition frequencies $\omega_{\rm X,1}\neq \omega_{\rm X,2}$ and $\omega_{\rm Y,1}\neq \omega_{\rm Y,2}$.  In this case each electronic transition of the system can be selectively addressed by choosing the right polarization and frequency of an external laser drive. In particular, the following two-photon driving Hamiltonian $H_{\rm }$ can be realized if two lasers of polarizations and amplitudes $\mathcal{E}_x \hat{x}$ and $\mathcal{E}_y \hat{y}$, and respective frequencies $\tilde{\omega}_{\rm X,1}=\omega_{\rm X,1}+\delta$ and $\tilde{\omega}_{\rm Y,2}=\omega_{\rm Y,2}-\delta$ are used to illuminate the system:
\begin{align}
    \frac{H_{\rm drive}}{\hbar}&=|g\rangle \langle x_{\rm S}|\sqrt{2}(\mathcal{E}_x e^{-{\rm i}\tilde{\omega}_{\rm X,1}t}+\mathcal{E}_x^\ast e^{{\rm i}\tilde{\omega}_{\rm X,1}t})\nonumber\\
    &+| x_{\rm S}\rangle \langle xy_{\rm S}|(\mathcal{E}_y e^{-{\rm i}\tilde{\omega}_{\rm Y,2}t}+\mathcal{E}_y^\ast e^{{\rm i}\tilde{\omega}_{\rm Y,2}t})\nonumber\\
    &+|g\rangle \langle y_{\rm S}|\sqrt{2}(\mathcal{E}_y e^{-{\rm i}\tilde{\omega}_{\rm Y,2}t}+\mathcal{E}_y^\ast e^{{\rm i}\tilde{\omega}_{\rm Y,2}t})\nonumber\\
    &+| y_{\rm S}\rangle \langle xy_{\rm S}|(\mathcal{E}_x e^{-{\rm i}\tilde{\omega}_{\rm X,1}t}+\mathcal{E}_x^\ast e^{{\rm i}\tilde{\omega}_{\rm X,1}t})\nonumber\\
    &+| x_{\rm S}\rangle \langle xx|\sqrt{2}(\mathcal{E}_x e^{-{\rm i}\tilde{\omega}_{\rm X,1}t}+\mathcal{E}_x^\ast e^{{\rm i}\tilde{\omega}_{\rm X,1}t})  \nonumber\\  
    &+| y_{\rm S}\rangle \langle yy|\sqrt{2}(\mathcal{E}_y e^{-{\rm i}\tilde{\omega}_{\rm Y,2}t}+\mathcal{E}_y^\ast e^{{\rm i}\tilde{\omega}_{\rm Y,2}t})+\text{H.c.}  \label{eq:drive}
\end{align}
If we further assume that $\delta<|\omega_{\rm X,1}-\omega_{\rm X,2}|,|\omega_{\rm Y,1}-\omega_{\rm Y,2}|$, the first two lines of Eq.\,\eqref{eq:drive} represent a drive that is nearly resonant with the respective electronic transitions, whereas the remaining lines are off resonant. Furthermore, the sum of the drive frequencies is resonant with the two-photon transition from the ground state $|g\rangle$ to the doubly excited state $|xy_{\rm S}\rangle$ ($\tilde{\omega}_{\rm X,1}+\tilde{\omega}_{\rm Y,2}=\omega_{\rm X,1}+\omega_{\rm Y,2}$). In this case it is possible to apply the rotating-wave approximation and  neglect the off-resonant terms:
\begin{align}
     \frac{H_{\rm drive}}{\hbar}&\approx|g\rangle \langle x_{\rm S}|\sqrt{2}\mathcal{E}_x^\ast e^{{\rm i}\tilde{\omega}_{\rm X,1}t}\nonumber\\
    &+| x_{\rm S}\rangle \langle xy_{\rm S}|\mathcal{E}_y^\ast e^{{\rm i}\tilde{\omega}_{\rm Y,2}t}+\text{H.c.}\label{eq:driveapprox}
\end{align}
%We finally apply the adiabatic approximation to derive the effective Hamiltonian of the driven system. To that end 
We derive the effective Hamiltonian of the driven system by first considering the dynamics of a trial wave function:
\begin{align}
    |\psi_{\rm drive}\rangle=a^{g}|g\rangle+a^{x_{\rm S}}|x_{\rm S}\rangle + +a^{y_{\rm S}}|y_{\rm S}\rangle + a^{xy_{\rm S}}|xy_{\rm S}\rangle,
\end{align}
under the Hamiltonian in Eq.\,\eqref{eq:driveapprox} expressed in the interaction picture with respect to the Hamitonian of the bare system (neglecting the small broadening due to spontaneous emission for the purpose of this derivation):
\begin{align}
    \frac{H_{\rm sys}}{\hbar}=\omega_{\rm X, 1}|x_{\rm S} \rangle\langle x_{\rm S} |+\omega_{\rm Y, 1}|y_{\rm S} \rangle\langle y_{\rm S} |+\omega_{xy_{\rm S}}|xy_{\rm S} \rangle\langle xy_{\rm S} |.
\end{align}
The following differential equations can be obtained:
\begin{align}
    \dot{a}^{g}=&-{\rm i}\sqrt{2}\mathcal{E}^\ast_{x}e^{{\rm i}\delta t}a^{x_{\rm S}}\nonumber\\
    &-{\rm i}\sqrt{2}\mathcal{E}^\ast_{y}e^{-{\rm i}(\omega_{\rm Y,1}-\tilde\omega_{\rm Y,2})t}a^{y_{\rm S}},\label{eq:ag}\\
    \dot{a}^{x_{\rm S}}=&-{\rm i}\sqrt{2}\mathcal{E}_{x}e^{{\rm i}(\omega_{\rm X,1}-\tilde\omega_{\rm X,1})t}a^{g}\nonumber\\
    &-{\rm i}\mathcal{E}^\ast_{y}e^{-{\rm i}(\omega_{\rm Y,2}-\tilde\omega_{\rm Y,2})t}a^{xy_{\rm S}},\label{eq:axs}\\
    \dot{a}^{y_{\rm S}}=&-{\rm i}\sqrt{2}\mathcal{E}_{y}e^{{\rm i}(\omega_{\rm Y,1}-\tilde\omega_{\rm Y,2})t}a^{g}\nonumber\\
    &-{\rm i}\mathcal{E}^\ast_{x}e^{-{\rm i}(\omega_{\rm X,2}-\tilde\omega_{\rm X,1})t}a^{xy_{\rm S}},\label{eq:ays}\\
    \dot{a}^{xy_{\rm S}}=&-{\rm i}\mathcal{E}_{x}e^{{\rm i}(\omega_{\rm X,2}-\tilde\omega_{\rm X,1})t}a^{y_{\rm S}}\nonumber\\
    &-{\rm i}\mathcal{E}_{y}e^{{\rm i}\delta t}a^{x_{\rm S}}.\label{eq:axys}
\end{align}
Equations\,\eqref{eq:axs} and \eqref{eq:ays} can be used to eliminate $a^{x_{\rm S}}$ and $a^{y_{\rm S}}$ in the adiabatic approximation:
\begin{align}
    a^{x_{\rm S}}&\approx \frac{\sqrt{2}\mathcal{E}_x a^g+\mathcal{E}^\ast_y a^{xy_{\rm S}}}{\delta} e ^{-{\rm i}\delta t}, \label{eq:axS}\\
    a^{y_{\rm S}}&\approx \frac{\sqrt{2}\mathcal{E}_y }{\tilde\omega_{\rm Y,2}-\omega_{\rm Y,1}}e^{-{\rm i}(\tilde\omega_{\rm Y,2}-\omega_{\rm Y,1})t}a^g\nonumber\\
    &+\frac{\mathcal{E}^\ast_x}{\omega_{\rm X, 2}-\tilde\omega_{\rm X,1}}e^{-{\rm i}(\omega_{\rm X, 2}-\tilde\omega_{\rm X,1})t} a^{xy_{\rm S}}.\label{eq:ayS}
\end{align}
Eqs. \eqref{eq:axS} and \eqref{eq:ayS} can be inserted into Eqs.\,\eqref{eq:ag} and \eqref{eq:axys}. Neglecting rotating terms and small energy shifts, the effective dynamics are
\begin{align}
    \dot{a}^g&=-{\rm i}g_{\rm eff} a^{xy_{\rm S}},\\
    \dot{a}^{xy_{\rm S}}&=-{\rm i}g^\ast_{\rm eff} a^{g},
\end{align}
which correspond to the effective Hamiltonian
\begin{align}
    H_{\rm drive}^{\rm eff}\approx \hbar g_{\rm eff}|g\rangle \langle xy_{\rm S}|+\text{H.c.},
\end{align}
with
\begin{align}
    g_{\rm eff}=\frac{\sqrt{2}\mathcal{E}^\ast_x\mathcal{E}_y^\ast}{\delta}.
\end{align}
This Hamiltonian induces Rabi oscillations between $|g\rangle$ and $|xy_{\rm S}\rangle$ with frequency $2 |g_{\rm eff}|$. If the illumination is applied for time $\tau_{\rm drive}=\pi/(2|g_{\rm eff}|)$ the system is driven from the ground state to the desired state $|xy_{\rm S}\rangle$. An analogous scheme exploiting the state $|y_{\rm S}\rangle$ with two lasers of polarizations and amplitudes $\mathcal{E}_x \hat{x}$ and $\mathcal{E}_y \hat{y}$, and respective frequencies $\tilde{\omega}_{\rm X,2}=\omega_{\rm X,2}-\delta$ and $\tilde{\omega}_{\rm Y,1}=\omega_{\rm Y,1}+\delta$ could be used instead.

\section{Weisskopf-Wigner approximation} \label{sec:wwapproximation}
Here we explicitly show how we obtain Eq. \eqref{eq:sscoefficient}, the wave function coefficient of the steady state electron-photon state. 
We reproduce the \textit{ansatz} for a general electron-photon wave function from Eq. \eqref{eq:generalwfn}:
\begin{multline}
    |\Psi(t)\rangle = \sum_{jk} c_{jk}^g|g\rangle a_{jX}^\dagger a_{kY}^\dagger |{\rm vac}\rangle + \sum_{j} c_{j}^{x_{\rm S}}|x_{\rm S}\rangle a_{jY}^\dagger|{\rm vac}\rangle \\ + \sum_{j} c_{j}^{y_{\rm S}}|y_{\rm S}\rangle a_{jX}^\dagger |{\rm vac}\rangle + c^{xy_{\rm S}} |xy_{\rm S}\rangle|{\rm vac}\rangle.
\end{multline}

The interaction Hamiltonian is:
\begin{align}
    H_{\rm int}&=\sum_j \Omega_{y_{\rm S},xy_{\rm S}} |y_{\rm S}, 1_{jX}, 0_{kY} \rangle\langle xy_{\rm S} |+\text{H.c.}\nonumber\\
    &+\sum_j \Omega_{x_{\rm S},xy_{\rm S}} |x_{\rm S}, 0_{kX}, 1_{jY} \rangle\langle xy_{\rm S} |+\text{H.c.}\nonumber\\
    &+\sum_{jk} \Omega_{g,y_{\rm S} } |g, 1_{jX}, 1_{kY}\rangle \langle y_{\rm S}, 1_{jX}, 0_{kY} |+\text{H.c.}\nonumber\\
    &+\sum_{jk} \Omega_{g,x_{\rm S}} |g, 1_{jX}, 1_{kY}\rangle \langle x_{\rm S}, 0_{jX}, 1_{kY} |+\text{H.c.}\nonumber\\
\end{align}

We now plug this state vector into the Schr\"odinger equation to derive the differential equations for the coefficients:
\begin{align}
    \frac{{\rm d}}{{\rm d}t}c^{xy_{\rm S}}&=-{\rm i}\omega_{xy_{\rm S}}c^{xy_{\rm S}}-{\rm i}\sum_j \Omega_{y_{\rm S},xy_{\rm S}} c^{y_{\rm S}}_{j}-{\rm i}\sum_j \Omega_{x_{\rm S},xy_{\rm S}} c^{x_{\rm S}}_{j},\label{eq:t_de1}\\
    \frac{{\rm d}}{{\rm d}t}c^{x_{\rm S}}_{j}&=-{\rm i}(\omega_{x_{\rm S}}+\omega_j)c^{x_{\rm S}}_{j}-{\rm i}\Omega_{x_{\rm S},xy_{\rm S}} c^{xy_{\rm S}}-{\rm i}\sum_k \Omega_{g,x_{\rm S}} c^{g}_{jk},\label{eq:t_de2}\\
    \frac{{\rm d}}{{\rm d}t}c^{y_{\rm S}}_{j}&=-{\rm i}(\omega_{y_{\rm S}}+\omega_j)c^{y_{\rm S}}_{j}-{\rm i}\Omega_{y_{\rm S},xy_{\rm S}} c^{xy_{\rm S}}-{\rm i}\sum_k \Omega_{g,y_{\rm S} } c^{g}_{jk},\label{eq:t_de3}\\
    \frac{{\rm d}}{{\rm d}t}c^{g}_{jk}&=-{\rm i}(\omega_{j}+\omega_k)c^{g}_{jk}-{\rm i} \Omega_{g,y_{\rm S} } c^{y_{\rm S}}_{j}-{\rm i}\Omega_{g,x_{\rm S}} c^{x_{\rm S}}_j,\label{eq:t_de4}
\end{align}
where we assume $\Omega_{op}$ is real. We now solve the differential equations in the Weisskopf-Wigner approximation. We first take Eq.\,\eqref{eq:t_de2} and formally integrate it:
\begin{align}
    c_j^{x_{\rm S}} &= c_j^{x_{\rm S}}(0) e^{-{\rm i}(\omega_{x_{\rm S}}+\omega_k)t}\nonumber\\
    &-{\rm i}\Omega_{x_{\rm S},xy_{\rm S}}\int_0^t e^{-{\rm i}(\omega_{x_{\rm S}}+\omega_k)(t-\tau)}c^{xy_{\rm S}}(\tau) {\rm d}\tau\nonumber \\
    &-{\rm i}\Omega_{g,x_{\rm S}}\int_0^t e^{-{\rm i}(\omega_{x_{\rm S}}+\omega_k)(t-\tau)} c^{g}_{jk}(\tau) {\rm d}\tau.
\end{align}
\begin{widetext}
We get an analogous equation for $c_j^{y_{\rm S}}$ and insert both into Eq.\,\eqref{eq:t_de1}:
\begin{align} \label{eq:t_effcxy}
    \frac{{\rm d}}{{\rm d}t}c^{xy_{\rm S}}=-{\rm i}\omega_{xy_{\rm S}}c^{xy_{\rm S}} -{\rm i}\sum_j \Omega_{y_{\rm S},xy_{\rm S}}\bigg( -{\rm i}\Omega_{y_{\rm S},xy_{\rm S}}\int_0^t e^{-{\rm i}(\omega_{y_{\rm S}}+\omega_k)(t-\tau)}c^{xy_{\rm S}}(\tau) {\rm d}\tau  -{\rm i}\Omega_{g,y_{\rm S}}\int_0^t e^{-{\rm i}(\omega_{y_{\rm S}}+\omega_k)(t-\tau)} c^{ g}_{jk}(\tau) {\rm d}\tau\bigg)\nonumber \\
    -{\rm i}\sum_j \Omega_{x,xy_{\rm S}}\bigg( -{\rm i}\Omega_{x_{\rm S},xy_{\rm S}}\int_0^t e^{-{\rm i}(\omega_{x_{\rm S}}+\omega_k)(t-\tau)}c^{xy_{\rm S}}(\tau) {\rm d}\tau
    -{\rm i}\Omega_{g,x_{\rm S}}\int_0^t e^{-{\rm i}(\omega_{x_{\rm S}}+\omega_k)(t-\tau)} c^{ g}_{jk}(\tau) {\rm d}\tau \bigg),
\end{align}
In the Weisskopf-Wigner approximation it is commonly assumed that the time integrals can be extended to infinity and that the $\tau$ dependent coefficients can be extracted from the integral by setting $\tau= t$. Since we are operating in the Schr\"odinger picture we have to perform this procedure with caution and we  have to define the slowly-varying amplitudes of a coefficient $c^{A}(\tau)=e^{-{\rm i}\omega_A \tau}\tilde c^{A}(\tau)$. We then set $\tilde c^A(\tau)\approx\tilde c^A(t)$, which is equivalent to performing the Markov approximation in the interaction picture. In this approximation we get:
\begin{align}
    -\sum_j |\Omega_{y_{\rm S},xy_{\rm S}}|^2\int_0^t e^{-{\rm i}(\omega_{y_{\rm S}}+\omega_k)(t-\tau)}e^{-{\rm i}\omega_{xy_{\rm S}}\tau}\,\tilde c^{xy_{\rm S}}(\tau) {\rm d}\tau \approx-\sum_j |\Omega_{y_{\rm S},xy_{\rm S}}|^2\tilde c^{xy_{\rm S}}(t)\int_0^t e^{-{\rm i}(\omega_{y_{\rm S}}+\omega_k)(t-\tau)}e^{-{\rm i}\omega_{xy_{\rm S}}\tau} {\rm d}\tau.
\end{align}
The integral in the last line can be further decomposed and the lower integration limit can be extended to $-\infty$:
\begin{align}
    &e^{-{\rm i}(\omega_{y_{\rm S}}+\omega_k)t}\int_{-\infty}^t e^{-{\rm i}(\omega_{xy_{\rm S}}-\omega_{y_{\rm S}}-\omega_j)\tau}{\rm d}\tau \approx e^{-{\rm i}\omega_{xy_{\rm S}}t}\left( \pi\delta(\omega_{xy_{\rm S}}-\omega_{y_{\rm S}}-\omega_j)+{\rm i}{\rm P}\left\{\frac{1}{\omega_{xy_{\rm S}}-\omega_{y_{\rm S}}-\omega_j}\right\} \right).
\end{align}
\end{widetext}
We further neglect the imaginary part of the parenthesis on the second line, the principal part (${\rm P}\{\}$) that generally leads to a spectral shift, and we retain only the delta function. We note that in the discrete case $\delta(\omega_k-\omega_j)\to \delta_{jk}/\Delta$ (which is a discrete representation of the delta function). Notice also that $e^{-{\rm i}\omega_{xy_{\rm S}}t}\tilde c^{xy_{\rm S}}(t)=c^{xy_{\rm S}}(t)$. We therefore get the result:
\begin{align}
    -\sum_j |\Omega_{y_{\rm S},xy_{\rm S}}|^2\int_0^t e^{-{\rm i}(\omega_{y_{\rm S}}+\omega_k)(t-\tau)} c^{xy_{\rm S}}(\tau) {\rm d}\tau\nonumber\\
    \approx- \frac{\pi |\Omega_{y_{\rm S},xy_{\rm S}}|^2}{\Delta} c^{xy_{\rm S}}(t)\equiv -\gamma_{y_{\rm S},xy_{\rm S}} c^{xy_{\rm S}}(t).
\end{align}
We get a similar result for the first term in the second parenthesis of Eq.\,\eqref{eq:t_effcxy}:
\begin{align}
    \approx-\gamma_{x_{\rm S},xy_{\rm S}} c^{xy_{\rm S}}(t).
\end{align}
The remaining terms in Eq.\,\eqref{eq:t_effcxy} yield after applying the same procedure:
\begin{multline}
    -\pi \sum_j [ \Omega_{y_{\rm S},xy_{\rm S}} \Omega_{g,y_{\rm S}} c_{jk}^{ g}(t)\delta(\omega_k-\omega_{y_{\rm S}})\\
    +\Omega_{x_{\rm S},xy_{\rm S}} \Omega_{g,x_{\rm S}} c_{kj}^{ g}(t)\delta(\omega_k-\omega_{x_{\rm S}})].
\end{multline}
This term is neglected in the calculations because of the frequency restriction imposed by the delta function, although in principle this term is of the same order as the terms leading to decay. We therefore obtain:
\begin{align}
 \frac{{\rm d}}{{\rm d}t}c^{xy_{\rm S}}=-{\rm i}\omega_{xy_{\rm S}}c^{xy_{\rm S}}-(\gamma_{x_{\rm S},xy_{\rm S}}+\gamma_{y_{\rm S},xy_{\rm S}}) c^{xy_{\rm S}}.  
\end{align}

Similarly we can derive the remaining differential equations:
\begin{align}
    \frac{{\rm d}}{{\rm d}t}c^{x_{\rm S}}_j &=-{\rm i}(\omega_{x_{\rm S}}+\omega_j)c_j^{x_{\rm S}}-\gamma_{g,x_{\rm S}} c^{x_{\rm S}}_j-{\rm i}\Omega_{x_{\rm S},xy_{\rm S}} c^{xy_{\rm S}},\\
    \frac{{\rm d}}{{\rm d}t}c^{y_{\rm S}}_j &=-{\rm i}(\omega_{y_{\rm S}}+\omega_j)c_j^{y_{\rm S}}-\gamma_{g,y_{\rm S}} c^{y_{\rm S}}_j-{\rm i}\Omega_{y_{\rm S},xy_{\rm S}} c^{xy_{\rm S}},\\
    \frac{{\rm d}}{{\rm d}t}c^{ g}_{jk}&=-{\rm i}(\omega_{j}+\omega_k)c^{ g}_{jk}-{\rm i} \Omega_{g,y_{\rm S}} c_j^{y_{\rm S}}-{\rm i} \Omega_{g,x_{\rm S}} c_k^{x_{\rm S}}.
\end{align}
This system of equations can be solved with the initial conditions:
\begin{align*}
    c^{xy_{\rm S}}(0)&=1,\\
    c_j^{x_{\rm S}}(0)&=c_j^{y_{\rm S}}(0)=c_{jk}^{ g}(0)=0,
\end{align*}
with the following steady-state solution in the rotating frame:
\begin{align}
\tilde{c}_{jk}^g(\infty)=\frac{\frac{-\Omega_{g,x_{\rm S} }\Omega_{x_{\rm S} ,xy_{\rm S} }}{i\omega_{x_{\rm S} }-i\omega_j+\gamma_{g,x_{\rm S} }}+\frac{-\Omega_{g,y_{\rm S}  }\Omega_{y_{\rm S} ,xy_{\rm S} }}{i\omega_{y_{\rm S} }-i\omega_k+\gamma_{g,y_{\rm S}  }}}{i(\omega_{xy_{\rm S} }-\omega_j-\omega_k)+\gamma_{x_{\rm S} ,xy_{\rm S} }+\gamma_{y_{\rm S} ,xy_{\rm S} }},
\end{align}
which matches Eq. \eqref{eq:sscoefficient}.

\section{Robust entanglement} \label{sec:robustentanglement}
The entanglement of the emitted photon pair is robust to changes in in $\omega_{X,1}$ relative to $\omega_{Y,1}$.

\begin{figure}[!htbp]
\centering
\includegraphics[width=1.0\linewidth]{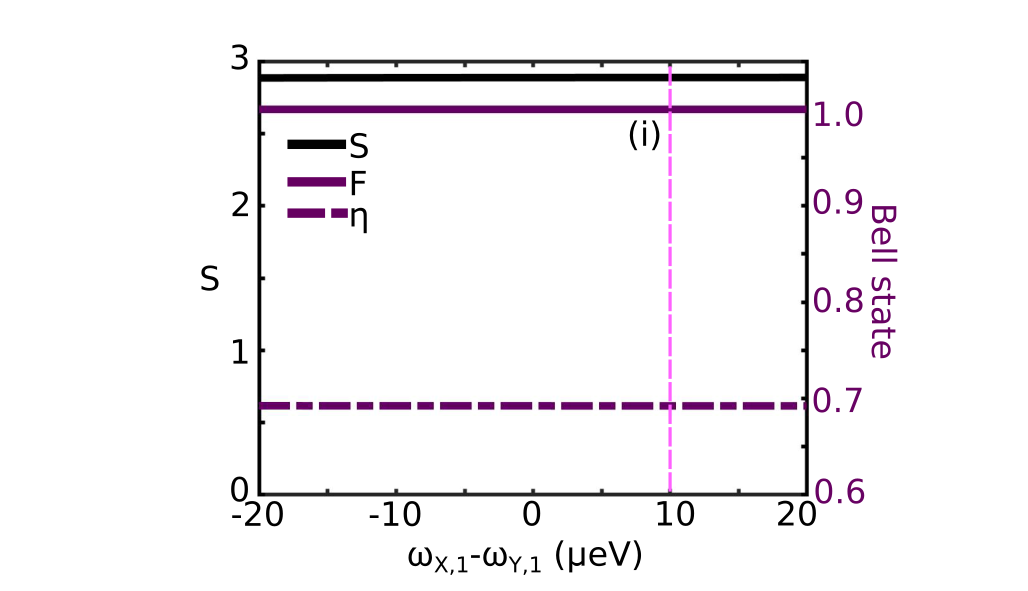}
\caption{Entanglement entropy $S$, Bell state fidelity $\mathcal{F}$, and Bell state efficiency $\eta$ are unaffected by varying $\omega_{X,1}$. The pink line (i) corresponds to the conditions in Fig. \ref{fig:ideal}.}
\label{fig:wxSwyS}
\end{figure}

\bibliography{biblio}

\end{document}